\documentclass[12pt]{article}
\usepackage{amsmath}

\title{Comment on ``The Free Will Theorem''}
\author{
Roderich Tumulka\footnote{Mathematisches Institut,
     Eberhard-Karls-Universit\"at, Auf der Morgenstelle 10, 72076
     T\"ubingen, Germany. E-mail:
     tumulka@everest.mathematik.uni-tuebingen.de}
}
\date{December 1, 2006}

\begin{document}
\maketitle
\begin{abstract}
In a recent paper \cite{CK05}, Conway and Kochen claim to have established that theories of the GRW type, i.e., of spontaneous wave function collapse, cannot be made relativistic. On the other hand, relativistic GRW-type theories have already been presented, in my recent paper \cite{Tum04} and by Dowker and Henson \cite{Fay02}. Here, I elucidate why these are not excluded by the arguments of Conway and Kochen.

\bigskip

 Key words: quantum theory without observers;
 Conway--Kochen ``free will'' theorem;
 Ghirardi--Rimini--Weber (GRW) theory of spontaneous wave function collapse;
 nonlocality.
\end{abstract}

\section{BACKGROUND} 
Physicists have often expressed the wish for a quantum theory without observers, i.e., a formulation of quantum theory that is not fundamentally about what observers will see if they carry out certain experiments, but instead about an objective reality. In such a theory, the quantum formalism gets derived (from the laws governing the objective reality) rather than postulated. One quantum theory without observers was presented by Ghirardi, Rimini, and Weber (hereafter, GRW) \cite{GRW86}. In its original version, it deals with non-relativistic quantum mechanics. In the GRW theory, the collapse of the wave function is an objective physical process; the rule for when and how the wave function collapses makes no reference to observers, but is expressed in a mathematical way as a stochastic process, thus describing a \emph{spontaneous collapse}. That is why observers are not mentioned in the definition of the theory, but can be treated as just another agglomeration of electrons and quarks. The objective reality in the GRW theory consists of the wave function together with certain variables describing the distribution of matter in space-time, called the \emph{primitive ontology} \cite{AGTZ06}. I will describe an example of primitive ontology later, the flash ontology.

\section{IMPOSSIBILITY CLAIM}
Conway and Kochen \cite{CK05} have argued that their ``free will theorem'' (hereafter, FWT) implies that a relativistic version of the GRW theory is impossible, as long as we insist (as I think we should) that experimenters can choose the setup of their experiments (such as the orientation of Stern--Gerlach magnets) at free will. This impossibility claim is inadequate since there exists an explicit counterexample \cite{Tum04}, a relativistic version of the GRW theory with flash ontology (hereafter, rGRWf). The lattice model of Dowker and Henson \cite{Fay02} (hereafter, DH) is another counterexample: a collapse theory on a discrete space-time that is relativistic in the appropriate lattice sense. Using rGRWf, I will point out three flaws in Conway and Kochen's argument. They admit \cite[Endnote 8]{CK05} that they ``have not made a detailed examination of this theory''; they should have.

\section{THE FREE WILL THEOREM} 
The FWT, which is not a mathematical theorem but a physical statement, is based on the consideration of an entangled pair of quantum particles, as in Einstein--Podolsky--Rosen (EPR) experiments. Writing ``freedom'' for the assumption that experimenters are free, the FWT asserts that
\begin{equation}\label{fwt}
  \text{freedom} \:+\: \text{QF} \:+\: \text{locality} \:+\: \text{determinism} 
  \:\Rightarrow\: \text{contradiction.}
\end{equation}
Here, QF stands for ``quantum formalism'' and means that the probabilities of outcomes of EPR-type experiments predicted by means of the usual quantum-theoretical rules are correct (approximately, at least). Locality means that events in a space-time region $A$ cannot influence other events in a space-time region $B$ if $A$ and $B$ are spacelike separated. Determinism means that the outcomes of experiments are functions of the past.

As a consequence of \eqref{fwt}, we have to abandon one of the four incompatible premises. It seems to me that any theory violating the freedom assumption invokes a conspiracy and should therefore be regarded as unsatisfactory (I will say a bit more about this in Sec.~\ref{sec:freedom}; see also \cite{Bell77}). The quantum probabilities have often been confirmed in experiment. This leaves us with giving up either locality or determinism.

Conway and Kochen argue that Lorentz invariance, together with the ``causality principle'' (``effects cannot happen at an earlier time than their causes''), implies locality (and, therefore, determinism must fail). This is not a sound argument, as we will see later in Sec.~\ref{sec:who}. Moreover, as also pointed out by Bassi and Ghirardi \cite{BG06}, locality is known to be wrong, thanks to Bell's theorem \cite{Bell64},
\begin{equation}\label{bt}
  \text{freedom} \:+\: \text{QF} \:+\: \text{locality} \:\Rightarrow\: \text{contradiction,}
\end{equation}
and Aspect's experiment \cite{aspect} confirming precisely those probability predictions used in Bell's argument; for discussions of Bell's theorem see, e.g., \cite{Bell87b, timbook, bbt, Nor06}. Thus, the option that Conway and Kochen recommend, to renounce determinism in order to save locality, does not, in fact, exist. This is the first flaw that I see.\footnote{How could such outstanding scientists make such a blatant mistake? Because they uncritically accepted a widespread but inadequate understanding of Bell's theorem, according to which the upshot of Bell's argument is that either locality or hidden variables have to be abandoned. (``J.~S.~Bell ... and others ... produce[d] no-go theorems that dispose of the most plausible hidden variable theories'' \cite[Sec.\ 1]{CK05}.) On the basis of that understanding, and with a background in orthodox quantum mechanics, it seems natural to give up hidden variables and retain locality. But in fact, this is not possible, as Bell's argument leaves no other choice than the failure of locality. For a nice and detailed discussion of exactly this point, see \cite{Nor06}.}

A certain irony, then, lies in the following implication shown already in 1935 by Einstein, Podolsky, and Rosen \cite{EPR35}:
\begin{equation}\label{epr}
  \text{freedom} \:+\: \text{QF} \:+\: \text{locality} \:\Rightarrow\: \text{determinism.}
\end{equation}
For \eqref{fwt} this means that abandoning determinism does not even resolve the contradiction. In fact, EPR \eqref{epr} and FWT \eqref{fwt} together show that, if freedom and QF are granted, locality must fail. That is why, as Bassi and Ghirardi have remarked, the FWT can be regarded as a proof of quantum nonlocality.

\section{RANDOMNESS}\label{sec:stoch}
The original non-relativistic GRW theory, the DH theory, and rGRWf: all of them are nonlocal and stochastic. They thus violate \emph{two} of the incompatible assumptions in \eqref{fwt}, locality and determinism. It would thus seem that Conway and Kochen, even with their erroneous view that locality can and should be saved at the expense of giving up determinism, should regard GRW-type theories as consistent with the FWT. They do not because they think that every stochastic theory is equivalent to a deterministic one, as far as the FWT is concerned. They must have in mind that any adequate theory is somehow neither deterministic nor stochastic, but belongs to a third class. I have little idea what that third class could be; they seem to think of ``particles with free will,'' whatever that means.

Conway and Kochen explicitly address the equivalence between stochasticity and determinism in a section entitled ``randomness can't help'' \cite[Sec.\ 10.1]{CK05}: 

\begin{quotation}
...let the stochastic element in a putatively relativistic GRW theory be a sequence of random numbers (not all of which need be used by both particles). Although these might only be generated as needed, it will plainly make no difference to let them be given in advance.
\end{quotation}

This is wrong, and that is the second flaw in their impossibility argument against GRW. To understand why it is wrong, it is best to look at how things play out in rGRWf.

\section{rGRWf}
This theory was defined in \cite{Tum04}; it is based on ideas of Bell \cite{Bell87}; for an easy introduction see \cite{Tum06}; for further discussion see \cite{AGTZ06, Mau05, Mau06}. It uses the \emph{flash ontology}, which means that it leads to a discrete set of space-time points, called flashes. That is, there are only finitely many flashes, at least in every finite space-time volume. The flashes replace the particle trajectories of classical mechanics: they are what matter consists of. ``A piece of matter then is a galaxy of such events'' \cite{Bell87}. For example, for some reasonable choice of the constants of nature of this theory, in a cubic centimeter of water there take place about $10^8$ flashes per second. For further discussion of the flash ontology and other possibly choices for the primitive ontology, see \cite{AGTZ06}.

The objective reality, according to this theory, consists of two things: the flashes (a set of space-time points), and the wave function (more precisely, a wave function on every spacelike hypersurface).\footnote{These two things are not independent of each other, they obey a functional relation: Given the flashes and the initial wave function $\psi_0$ on some spacelike hypersurface $\Sigma_0$, one obtains the wave function $\psi_\Sigma$ on any other spacelike hypersurface $\Sigma$ by choosing an arbitrary spacelike foliation $t \mapsto \Sigma_t$ connecting $\Sigma_0$ to $\Sigma$, applying the Dirac evolution and collapsing the wave function in the position representation (see \cite{Tum04,Tum06} for the precise equation) whenever the hypersurface crosses a flash; the result $\psi_\Sigma$ does not depend on the choice of the interpolating hypersurfaces $\Sigma_t$. Conversely, given the wave function on at least all hypersurfaces $\Sigma_t$ belonging to one spacelike foliation of space-time, we get back the flashes as the space-time points where the wave function collapses are centered.} The set of flashes is random (it is a point process in space-time), and its Lorentz-covariant distribution is defined by the (initial) wave function and one (initial) flash for every ``particle type'', by means of a formula that was written down explicitly in \cite{Tum04,Tum06}.

\section{EPR EXPERIMENTS IN rGRWf}\label{sec:epr}
The rGRWf theory has been defined so far only for a system of $N$ \emph{non-interacting} quantum particles.\footnote{The DH theory, however, allows for interaction.} It would certainly be nice to have a version that incorporates interaction, but for EPR experiments this is unnecessary: We can treat the situation as a problem involving only two particles, without interaction. Instead of including the experimenters in the theoretical treatment as large systems of interacting electrons and quarks, we just include (say) the Stern--Gerlach magnetic field that the experimenter arranges as the external magnetic field. Instead of including the detectors as large systems, we just wait long enough (about $10^8$ years), until the wave function has spontaneously collapsed for the first time. After that, the result of any later detection is pre-determined by the collapsed wave function because it is nearly a position eigenstate, that is, because the collapsed wave function is almost entirely concentrated in only one of the channels in which the particle could be detected.

In view of the fact that interaction plays no essential role for EPR experiments, and thus for the FWT, it is somewhat inadequate that Conway and Kochen write: ``We ... cannot say which of our assumptions would fail in a valid extension of [rGRWf] by an interaction term'' \cite[Endnote 8]{CK05}. They should have written: ``we ... cannot say which of our assumptions fails in rGRWf.'' If a valid extension of rGRWf by an interaction term is a counterexample to their impossibility claim, then so is rGRWf without interaction.

Here is the answer which assumptions fail: Freedom is true in rGRWf in the sense that the theory provides, for any given the external fields, a distribution of the flashes. QF is true in the sense that the outcome of the detections are pre-determined by the first flash on each side, and that the distribution of these flashes agrees (very very closely) with the probabilities prescribed by the quantum formalism. As mentioned before, locality and determinism are violated.

For our EPR experiment, consider two spacelike separated regions $A$ and $B$ of space-time (of sufficient timelike extension, $10^8$ years, for our purposes), and two entangled but non-interacting particles, one in $A$ and one in $B$. I suppose the external (Stern--Gerlach) fields, $F_A$ and $F_B$, are such that each particle can end up in one out of two or more channels.

Now, what is an EPR experiment like in rGRWf? There is a random set of flashes in space-time, some in region $A$ and some in region $B$. (Leave aside all the flashes occurring before the wave packet passes the Stern--Gerlach magnet.) Let us look at the flashes from the viewpoint of an arbitrary Lorentz frame with hypersurfaces $\Sigma_t$; then one of the flashes occurs first. The probability for the first flash occurring in $A$ is $\frac{1}{2}$, and the same for $B$. Suppose that the first flash occurs in $A$, and call it $f_A$; it occurs in one of the channels behind the Stern--Gerlach magnet of $A$ (because it does not occur where the wave function vanishes). The probability for $f_A$ occurring in the  $\alpha$-th channel agrees (very closely) with the (marginal) probability prescribed by the quantum formalism for the outcome on $A$'s side being $\alpha$. When $f_A$ occurs, the wave function on the hypersurfaces $\Sigma_t$ collapses in very much the same way it would collapse in an orthodox treatment as soon as the $A$ particle is detected in the channel  $\alpha$: all parts of the wave function corresponding to other channels for the $A$ particle are set to (very nearly) zero. Now consider the second flash. Again, it has probability $\frac12$ to occur in region $A$, in which case it also occurs in the  $\alpha$-th channel and does not change the wave function in an essential way. Let us wait until the first flash $f_B$ occurs in region $B$. It occurs in one of the $B$ channels. The probability that $f_B$ occurs in the  $\beta$-th channel agrees (very closely) with the (conditional) probability prescribed by the quantum formalism for the outcome on $B$'s side being  $\beta$, \emph{given that the outcome on $A$'s side was  $\alpha$}. Any further flash in the region $B$ will again occur in the  $\beta$-th channel.

As a consequence, the joint distribution of the flashes agrees with the joint distribution of the outcomes prescribed by the quantum formalism, as I said above.

\section{WHO INFLUENCES WHOM}\label{sec:who}
Another consequence is that the (conditional) distribution of $f_B$ depends on the location  $\alpha$ of $f_A$. This is where locality is violated: $f_A$ influences $f_B$, despite the spacelike separation. But the direction of the influence depends on the Lorentz frame: In another frame, in which $f_B$ precedes $f_A$, the distribution of $f_B$ is the (marginal) quantum probability for outcome  $\beta$, and the conditional distribution of $f_A$, given that $f_B$ occurred in channel  $\beta$, is the conditional quantum distribution of outcome  $\alpha$ given that the outcome on $B$'s side was  $\beta$; this is simply a consequence of probability calculus and of conditionalizing on the past. Thus, in this other frame, $f_B$ influences $f_A$. Who influences whom is frame dependent. There is no objective fact about who ``really'' influenced whom. There is no need for such a fact. The objective facts are where-when the flashes occur, and it is enough if a theory prescribes, as does rGRWf, their joint distribution in a Lorentz-invariant way. Whether nature chooses the space-time point $f_B$ first, and $f_A$ afterwards, or the other way around, does not seem like a meaningful question to me.

Indeed, this situation is very similar to Conway and Kochen's thoughts about influences in a relativistic space-time. Their toy model of a ``hexagonal universe'' \cite[Sec.\ 4.1]{CK05} is a nice illustration of a frame-dependent direction of influence. However, from the way they refer to the demon ``Janus,'' who makes all the random decisions but is obliged to make them in the ordering of some frame (thus breaking Lorentz invariance), it is not clear to me whether they have realized that a relativistic GRW theory is not obliged in this way.

Conway and Kochen have even introduced a word for the frame-dependence of influences: effective causality. (They call all influences ``causality,'' while I find that the word suggests that the influence has a direction, which is not present in reality. Whenever they say that ``$y$ is an effect of the cause $x$,'' they seem to mean (merely) that $x$ influences $y$; for simplicity, I will adopt this terminology in the following. Thus, the ``causality principle'' means that one event cannot influence an earlier one.) A theory is effectively causal if in each frame the causality principle holds. This is the case, for example, in rGRWf, because whether $f_A$ influences $f_B$ or vice versa is frame-dependent. That is, rGRWf is effectively causal despite the nonlocality. By the way, this undermines Conway and Kochen's reason for assuming locality: Effective causality (i.e., the ``causality principle'' in every frame) does not, together with Lorentz invariance, imply locality, as rGRWf shows.

\section{RANDOMNESS VERSUS DETERMINISM}\label{sec:freedom}
Now that we have seen how rGRWf behaves in an EPR experiment, could we make it deterministic, following the recipe that Conway and Kochen suggest in the passage I quoted in Sec.~\ref{sec:stoch}? Since the random element in rGRWf is the set of flashes, nature should, according to the recipe, make at the initial time the decision where-when flashes will occur, make this decision ``available'' to every space-time location, and have the flashes just carry out the pre-decided plan. The problem is that the distribution of the flashes depends on the external fields, and thus on the free decision of the experimenters. In particular, the correlation between the flashes in $A$ and those in $B$ depends on \emph{both} external fields. Thus, to let the randomness ``be given in advance'' would make a big difference indeed, as it would require nature to know in advance the decision of both experimenters, and would thus require the theory either to give up freedom or to allow influences to the past.

Let me approach the same point from a different, more general, perspective: A crucial difference between stochastic and deterministic theories is that, if freedom is granted, a stochastic theory can manage to be both effectively causal and nonlocal, but a deterministic one cannot. To see this, note first that rGRWf is not only effectively causal concerning the influences between flashes, but also concerning the influences from the external fields $F_A$, $F_B$ to the flashes: In the situation considered in Sec.~\ref{sec:epr}, the first flash $f_A$ does not depend on the field $F_B$, in a frame in which the points of $B$ are later that those of $A$. Now, concerning deterministic theories, suppose that, following a variant of the recipe, the initial randomness is just a coin tossing sequence $X_1,X_2,\ldots$---random bits---and there is a function $f_y(X_1,X_2,\ldots)$ that determines whether or not there is a flash at space-time point $y$.\footnote{The validity of the reasoning that follows is, in fact, not limited to the flash ontology, but holds for every choice of ontology localized at a space-time point $y$, in particular for particle world lines and fields.} We cannot expect that the same function $f_y$ works for every choice of external fields. So let us write $f_y = f_y(F_A,F_B,X_1,X_2, \ldots)$. A collection of functions $f_y$, one for each $y$, then represents a deterministic theory. It is now relevant that the choice of the $f_y$ is not unique. (For example, starting from rGRWf, every frame $\Lambda$ provides a different such choice, $f_y = f_y^\Lambda$, by making all the random decisions in the temporal ordering of events given by $\Lambda$---like Janus, but using the random bits. Then, in this frame $\Lambda$, the function $f_y^\Lambda$ does not depend on the future magnetic fields, i.e., on the field at space-time points which in this frame are later than $y$. However, $f_y^\Lambda$ will entail influences to the past in other frames!) Generally, for every given $f_y$ there is always a frame in which $f_y$ depends on the future, i.e., on the field at space-time points later than $y$. That is because otherwise $f_y$ could depend only on, besides the random bits, the fields inside the past light cone of $y$, and would hence be local. Hence, a deterministic theory cannot be both effectively causal and nonlocal. 

To be sure, there is a sense in which stochastic theories are indeed equivalent to deterministic ones: If all particles of the universe are included in the theoretical treatment, including the experimenters, then nature might as well make all the random decisions right at the initial time. But a different perspective is relevant: We should require a physical theory to be non-conspiratorial, which means here that it can cope with arbitrary choices of the experimenters, as if they had free will (no matter whether or not there exists ``genuine'' free will). A theory seems unsatisfactory if somehow the initial conditions of the universe are so contrived that EPR pairs always know in advance which magnetic fields the experimenters will choose. That is why the freedom assumption is relevant,\footnote{A side remark: On their face, freedom and determinism appear to contradict each other, since in a deterministic universe the experimenters cannot have genuine free will. But the lack of genuine free will is not relevant. What is relevant is that the theory provides a story about how the EPR pair reacts to \emph{any} external field, consistent with the probabilities given by the QF.} and why it is relevant to leave the experimenters out of the theoretical treatment. And \emph{that} is when stochastic theories are no longer equivalent to deterministic ones. That was the second flaw. Now I turn to the third.

\section{LOCALITY, EFFECTIVE LOCALITY,\\ NO-SIGNALLING}
Conway and Kochen fail to distinguish sharply enough between locality, no-signalling, and what could be called ``effective locality.'' The way they formulate their ``FIN axiom'', it is \emph{effective locality}, but the way they use it in deriving the contradiction, it is \emph{locality}, as I will show presently. Later, when they prove the consistency of their axioms, it is \emph{no-signalling} (``inhabitants ... cannot transmit information'' etc. \cite[Sec.\ 4.2]{CK05}).

What do these notions mean, and what is the difference? No-signalling means the impossibility for intelligent beings to transmit messages or signals faster than light. In rGRWf, for example, no-signalling holds \cite{Tum04,Tum06} but not locality.\footnote{The same is true of the non-relativistic GRW theory and of Bohmian mechanics, if understood in the appropriate sense.} This fact underlines the relevance to distinguish between locality and no-signalling. The terminology of ``effective'' properties was coined by Conway and Kochen. Here is their definition of \emph{effective causality}: ``the universe should appear causal from every inertial coordinate frame'' \cite[Sec.\ 6]{CK05}. And here is their definition of \emph{effective transmission of information}: ``If information is really transmitted from $a$ to $b$, then this will appear to be so in all coordinate frames, which we shall express by saying that information is effectively transmitted from $a$ to $b$'' \cite[Sec.\ 6]{CK05}. I take this to mean that if in \emph{some} frames no information is transmitted from $a$ to $b$, while in other frames perhaps some information is, then there is \emph{no effective} transmission of information. That such is the case for spacelike separated regions $a$ and $b$ is what I am calling effective locality, and is the wording of the FIN axiom. Locality, in contrast, means, in this terminology, that in \emph{every} frame no information is transmitted from $a$ to (spacelike separated) $b$. In rGRWf, for example, no effective transmission of information is involved in an EPR experiment, since it depends on the frame whether information is transmitted from $f_A$ to $f_B$ or vice versa, and no transmission takes place about which all frames agree. As a consequence, ironically, rGRWf actually \emph{satisfies} the FIN axiom as formulated in \cite[Sec.\ 1.1]{CK05}. Thus, given that rGRWf is not deterministic, it may even seem as if rGRWf was exactly the kind of theory that Conway and Kochen wanted to advocate.

Similarly, as mentioned before, rGRWf is effectively causal. In contrast, Bohmian mechanics for a preferred frame is neither effectively causal nor effectively local since if two spacelike separated events, $e$ and $f$, are such that $e$ precedes $f$ in the preferred frame, then $e$ influences $f$, and not vice versa, in all frames. Since Bohmian mechanics satisfies no-signalling, we see that the three concepts locality, effective locality, and no-signalling are really different. In the DH theory, the situation is qualitatively the same as in rGRWf: no-signalling holds, and DH is effectively local and effectively causal, but not local.

But let me show where locality is used in the derivation of the FWT. It is striking that the FIN axiom is nowhere mentioned by name. The passage where it is used, or rather where locality is used, reads: 
\begin{quotation}
Now we defined $\alpha'$ so as to be independent of $x$, $y$, $z$, but it is also independent of $w$, since there are coordinate frames in which B's experiment happens later than A's. Similarly, $\beta'$ is independent of of $x$, $y$, $z$ as well as $w$.
\end{quotation}
Here, $x$, $y$, $z$ correspond to the experimenter's choice (the external field $F_A$) on side $A$ and $w$ to that on side $B$; $\alpha'$ ($\beta'$) is all the information on which the outcome on side $A$ ($B$) may depend, except on $x$, $y$, $z$ (on $w$). The quoted passage means it is assumed in \emph{every} frame that $\alpha'$ is independent of $w$, \emph{and} that $\beta'$ is independent on $x$, $y$, $z$. That is locality, not effective locality. Note the logical gap: Whereas effective locality requires that in \emph{some} frames no transmission takes place, at this point in the proof it is taken for granted that in \emph{every} frame no transmission takes place. ``Some'' becomes ``every.'' This is the third flaw.

\section{CONCLUSION}

The three flaws in Conway and Kochen's impossibility argument against relativistic GRW theories were: the assumption of locality; the equivalence between stochastic and deterministic theories; and mixing up locality, effective locality, and no-signalling. Relativistic GRW theories are not excluded, and the rGRWf theory nicely illustrates how the obstructions can be overcome. Indeed, even deterministic theories are not excluded, as long as they are nonlocal.

\bigskip

\noindent\textit{Acknowledgments.} I thank Angelo Bassi (LMU M\"unchen), Detlef D\"urr (LMU M\"unchen), GianCarlo Ghirardi (Universit\`a di Trieste and ICTP), Shelly Goldstein (Rutgers University), Tim Maudlin (Rutgers University), and Nino Zangh\`\i\ (Universit\`a di Genova) for helpful discussions, and an anonymous referee for helpful remarks on a previous version of this article.

\end{document}